\documentclass[12pt]{article}

\input tcilatex
\QQQ{Language}{
American English
}

\begin{document}

\author{A.I.Volokitin$^{1,2}$ and B.N.J.Persson$^1$ \\
\\
$^1$Institut f\"ur Festk\"orperforschung, Forschungszentrum \\
J\"ulich, D-52425, Germany\\
$^2$Samara State Technical University, 443010 Samara,\\
Russia}
\title{Frictional drag between quantum wells mediated by fluctuating
electromagnetic field}
\maketitle

\begin{abstract}
We use the theory of the fluctuating electromagnetic field to calculate the
frictional drag between nearby two-and three dimensional electron systems.
The frictional drag results
from coupling
via a fluctuating electromagnetic
field, and can be considered as the dissipative part of the van der Waals
interaction.
In comparison with other similar calculations for semiconductor two-dimensional
system we include retardation effects. 
 We consider the dependence of the frictional drag force on the
temperature $T$, electron density and separation $d$. 
We find, that retardation effects become dominating factor for high electron
densities, corresponding thing metallic film
, and suggest a new experiment to
test the theory. The relation between friction and heat transfer is also
briefly commented on.
\end{abstract}

\section{ Introduction}

A great deal of attention has been recently devoted to double layer systems
in which two parallel quasi-two-dimensional (2D) subsystems (electron or
hole gases) are separated by a potential barrier thick enough to prevent
particles from tunneling across it but allowing for the interaction between
particles on both its sides. Some time ago, Pogrebinskii and later Price

\cite{Pogrebinskii} predicted that the Coulomb interaction between two 2D
electron systems will induce a frictional drag force between the layers: a
current in one film will induce a current in the adjacent film. The first
frictional drag experiment was performed by Gramila \textit{et al.} for two
electron layers \cite{Gramila1,Gramila2} and by Sivan, Solomon and Shtrikman
for an electron-hole system \cite{Sivan}. In these experiments a current is
drawn in the first layer, while the second layer is an open circuit. Thus no
dc-current can flow in the second layer, but an induced electric field occur
that opposes the `` drag force '' from the first layer. These experiments
spurred a large body of theoretical work both on electron-hole systems \cite
{electron-hole} and on electron-electron systems \cite{elel1}-\cite
{elel11} \ . Most of this work focused on interlayer Coulomb interaction,
the most obvious coupling mechanism and the one considered in the original
theoretical papers \cite{Pogrebinskii}, though the contributions due to an
exchange of phonons between the layers have also been considered \cite
{Gramila2,phonon1,phonon2,phonon3}. The origin of Coulomb drag is due to
quantum and thermal fluctuations of the charge and current densities and can
be considered as the dissipative part of van der Waals interaction. The
static aspects of van der Waals interaction is well understood, and from the
theory of Lifshitz \cite{Lifshitz} it is known that one must distinguish
between two distance regime : (a) The nonretarded limit, when the separation
between bodies $d$ is small compared to wavelengths $\lambda \sim c/\omega
_0 $, where $\omega _0$ is a characteristic frequency of the charge
fluctuation, and $c$ the light velocity, the interaction is determined by
the fluctuations in an instantaneous Coulomb field . For metal $\omega
_0\sim \omega _p$, where $\omega _p$ is the plasma frequency .(b)
Retardation effects become important, when $d>\lambda $ . As we have shown in
Ref. \cite{Volokitin,Persson1}, when calculating the dissipative part of the van der
Waals interaction for two semi-infinite bodies in relative motion,
retardation effects become important for $d>c/\omega _0$, where $\omega
_0\sim \omega _p(\omega_p\tau)$ and $\tau $ is the relaxation time.
For $\omega_p\sim 10^{16}s^{-1}$ and $\tau\sim 10^{-14}s$ retardation effects
become important for very short distances $d>1 \ {\rm \AA}$.
However for 2D systems there is no investigation of the role of retardation
effects in the frictional drag experiments. For large distances the retarded
contribution to the frictional drag becomes important, and it is interesting
to compare this contribution to the non-retarded contribution. To evaluate
the retarded contribution from photon exchange we use the general theory of
the fluctuating electromagnetic field developed by Rytov \cite{Rytov} and
applied by Lifshitz \cite{Lifshitz} for studying the conservative part, and
by us \cite{Volokitin} for studying dissipative part of the van der Waals
interaction. In this approach the interaction between the bodies is mediated
by the fluctuating electromagnetic field which is always present in the
vicinity of any collection of atoms. Beyond the boundaries of a solid this
field consist partly of traveling waves and partly of evanescent waves which
are damped exponentially with the distance away from the surface of the
body. The method we use for calculating the frictional drag between two
nearby 2D systems is quite general, and is applicable to any body at
arbitrary temperature. It takes into account retardation effects, which
become important for large enough separation between the bodies.

We shall calculate frictional stress $\sigma = \gamma v$ acting on the
electrons in
layer 1 due to the current density $J_2=n_2ev$ in the layer 2,
where $n_2$ is the carrier concentration (per unit area).
If no current is allowed to
flow in layer 1 (open circuit) an electric field $E_1$ develops whose influence
cancels the frictional stress $\sigma$
between the layers.
The frictional stress $\sigma = \gamma v$ must equal the
induced stress $n_1 e E_1$ so that
$$\gamma = n_1 eE_1 /v= n_1 n_2 e^2 E_1/J_2=n_1 n_2 e^2 \rho_{12},
$$
where
the \textit{transresistivity }$\rho _{12}=E_1/J_2$ is defined as the
ratio of the induced electric field in the first layer to the driving
current density in the second layer.
The transresistivity is often interpreted
in terms of a drag rate which, in analogy with a Drude model, is defined by $%
\tau _D^{-1}=\rho _{12}n_{2}e^2/m^{*}=\sigma /n_{1}m^{*}v.$

We find that for modulation-doped semiconductor
quantum wells retardation effects are not important under typical
experimental conditions, supporting earlier calculations where retardation
effects always
have been neglected \cite{electron-hole}, \cite{elel1}-\cite{elel11} . However, although 
previous calculation for friction drag between two-dimensional semiconductor systems
are equivalent to ours, other approaches were very different. The present derivation 
offers an alternative insight and is more general. A striking new result
we find, that for systems with high 2D-electron density,
e.g., thin metallic
films, retardation effects becomes crucial and in fact dominates the frictional
shear stress $\sigma$.
To test the theoretical predictions presented below we therefore suggest
performing experiments on
thin metallic layers grown on insulating
substrates and separated by thin insulating layers.
For example, for two thin (of order monolayer) silver films
separated by $d \sim 100 \ {\rm \AA}$,
we estimate that the induced voltage $U_1$ in metal film
1, due to a current $J_2$ in layer 2, will be of order $U_1\approx
10^{-8} \ U_2$, where $U_2$ is the driving
voltage applied to metal film 2. Thus if $U_2\approx 1 \ {\rm V}$ the induced
voltage will be of order
$10 \ {\rm pV}$ which should be possible to detect experimentally.
We note that the study of this problem is also
of direct interest in the context of sliding friction, since the electronic
friction
probed when two metallic bodies slide relative to each other, should be the
same as the
electronic friction probed by the transresistivity measurement, see Fig. 1. The
electronic
sliding friction (usually called vacuum friction) has recently been invoked to
explain
experimental results for
the damping of a small metal particle vibrating in the vicinity of a flat metal
surface \cite{Dorofeev} ,
but this explanation is controversial \cite{Persson1} , and it is
clear that independent studies of the electronic friction would be of great
interest.

\section{Calculation of the fluctuating electromagnetic field}

Let us firstly calculate the fluctuating electromagnetic field from one 2D
system, surrounded by a dispersionless dielectric medium. We introduce a
coordinate system with the $xy-$plane in the 2D layer. Following Lifshitz
\cite{Lifshitz}, to calculate the fluctuating field we shall use the general
theory due to Rytov, which is described in his book \cite{Rytov}. This
method is based on the introduction of a ``random '' field in the Maxwell
equations (just as, for example, one introduce a ``random'' force in the
theory of Brownian motion of a particle ).For a monochromatic field (time
factor $\exp (-\mathrm{i}\omega t)$) in a dielectric, nonmagnetic medium,
these equations are:
\begin{eqnarray}
\mathbf{\nabla \times E} &=&\limfunc{i}\frac \omega c\mathbf{B}  \nonumber \\
\mathbf{\nabla }\times \mathbf{H} &=&-\limfunc{i}\frac \omega c\mathbf{D+}%
\frac{4\pi }c(\mathbf{j+j}_f)\delta (z)  \label{one}
\end{eqnarray}
where, following to Rytov, we divided the total current density $\mathbf{j}%
_{tot}$ into two parts, $\mathbf{j}_{tot}=\mathbf{j}+\mathbf{j}_f$, the
fluctuating current density $\mathbf{j}_f$ associated with thermal and
quantum fluctuations, and the current density $\mathbf{j}_{\text{ }}$
induced by the electric field $\mathbf{E}$. $\mathbf{D,H}$ and$\mathbf{\,}$ $%
\mathbf{B}$ are the electric displacement field, the magnetic and the
magnetic induction fields, respectively. For nonmagnetic medium $\mathbf{B=H}
$ and $\mathbf{D}=\varepsilon \mathbf{E}$, where $\varepsilon $ is the
dielectric constant of the surrounded media. Eliminating $\mathbf{B}$ and $%
\mathbf{H}$ from (\ref{one}) one get
\begin{equation}
\mathbf{\nabla }^2\mathbf{E}+\left( \frac \omega c\right) ^2\varepsilon
\mathbf{E}-\frac{4\pi }\varepsilon \mathbf{\nabla (}\rho +\rho _f)\delta (z)+%
\frac{4\pi \limfunc{i}\omega }{c^2}(\mathbf{j+j}_f)\delta (z)=0  \label{two}
\end{equation}
where the total charge density $\rho +\rho _f$ is the sum of the induced and
fluctuating electron densities\textbf{.} We represent the current and
electron densities in the form of a Fourier series
\begin{equation}
\mathbf{j}(\mathbf{r)}=\frac 1{\sqrt{A}}\sum_{\mathbf{q}}\mathbf{j(q})e^{%
\limfunc{i}\mathbf{q\cdot r}}  \label{three}
\end{equation}
\begin{equation}
\rho (\mathbf{r)}=\frac 1{\sqrt{A}}\sum_{\mathbf{q}}\rho (\mathbf{q})e^{%
\limfunc{i}\mathbf{q\cdot r}}  \label{three1}
\end{equation}
where $\mathbf{q}$ and $\mathbf{r}$ are 2D vectors in the $xy$-plane and $A$
is the surface area. From the equation of continuity
\[
\frac{\partial \rho }{\partial t}+\mathbf{\nabla \cdot j\ =}0
\]
one get $\omega \rho =\mathbf{q\cdot j}.$ Then (\ref{two}) takes the form
\[
\ \frac{\limfunc{d}^2\mathbf{E}(\mathbf{q})}{\limfunc{d}^2z}+p^2\mathbf{E}(%
\mathbf{q})-\frac{4\pi i\mathbf{q}}{\varepsilon \omega }\mathbf{q\cdot }(%
\mathbf{j+j}_f^{})\delta (z)-\frac{4\pi \mathrm{\hat z}}{\varepsilon \omega }%
\mathbf{q\cdot }(\mathbf{j+j}_f)\delta ^{\prime }(z)
\]
\begin{equation}
+\frac{4\pi \limfunc{i}\omega }{c^2}(\mathbf{j+j}_f)\delta (z)=0
\label{four}
\end{equation}
where

\begin{equation}
p=\sqrt{\left( \frac \omega c\right) ^2\varepsilon -q^2}  \label{five}
\end{equation}
and \textrm{\^z} is a unit vector along the $z$ -axis. We shall consider
separately the two cases where the electric field $\mathbf{E}$ is in the
plane determined by the vectors $\mathrm{\hat q}=\mathbf{q/}q$ and \textrm{%
\^z} ($p-$ polarized waves) and perpendicular to this plane along the vector
$\mathbf{n}=\mathrm{\hat z}\times \mathrm{\hat q}$ ($s-$polarized waves).

Let us firstly suppose that $\mathbf{E}$ is parallel the vector $\mathbf{n}$%
. In this case (\ref{four}) gives
\begin{equation}
\frac{\limfunc{d}^2E_{\mathbf{n}}}{\limfunc{d}^2z}+p^2E_{\mathbf{n}}+\frac{%
4\pi \limfunc{i}\omega }{c^2}\left( \sigma _t(\mathbf{q},\omega )E_{\mathbf{n%
}}+j_{f\mathbf{n}}^{}\right) \delta (z)=0  \label{six}
\end{equation}
where we have used Ohm's law
\begin{equation}
j_{\mathbf{n}}=\sigma _t(\mathbf{q},\omega )E_{\mathbf{n}}  \label{seven}
\end{equation}
where $\sigma _t$ is the transverse conductivity of the layer\textbf{. }The
solution of (\ref{six}) can be written in the form
\begin{equation}
E_{\mathbf{n}}=u_{\mathbf{n}}e^{ip\left| z\right| }  \label{eight1}
\end{equation}
>From (\ref{six}) one can get the following boundary conditions at $z=0$
\begin{eqnarray}
E_{\mathbf{n}}(z &=&+0)=E_{\mathbf{n}}(z=-0)  \nonumber \\
\frac{\limfunc{d}E_{\mathbf{n}}}{\limfunc{d}z}\left| _{z=+0}\right. -\frac{%
\limfunc{d}E_{\mathbf{n}}}{\limfunc{d}z}\left| _{z=-0}\right. &=&-\frac{4\pi
\limfunc{i}\omega }{c^2}\left( \sigma _t(\mathbf{q},\omega )E_{\mathbf{n}%
}+j_{f\mathbf{n}}^{}\right)  \label{nine}
\end{eqnarray}
Substituting (\ref{eight1}) into boundary conditions (\ref{nine}) we get the
following relation between the fluctuating current density $j_{f\mathbf{n}}$
and electric field $E_{\mathbf{n}}$%
\begin{equation}
j_{f\mathbf{n}}=-\left( \sigma _t+\frac{pc^2}{2\pi \omega }\right) E_{%
\mathbf{n}}(z=0)  \label{ten}
\end{equation}

On the other hand the electric field can be calculated using linear response
theory \cite{Landau1}. The Hamiltonian of the system has the form $\hat
H+\hat H_{int}$, where $\hat H$ is the Hamiltonian of the body and radiation
field, while
\begin{equation}
\hat H_{int}=-\frac 1c\int \hat A_{\mathbf{n}}(\mathbf{r})j_{f\mathbf{n}}(%
\mathbf{r)}e^{-i\omega t}d^2\mathbf{r=}-\frac 1c\sum_{\mathbf{q}}\hat A_{%
\mathbf{n}}(-\mathbf{q})j_{f\mathbf{n}}(\mathbf{q,}\omega \mathbf{)}%
e^{-i\omega t}  \label{eleven}
\end{equation}
Accordingly to the linear response theory the average value of the vector
potential is determined by
\begin{equation}
\left\langle \hat A_{\mathbf{n}}(\mathbf{q,\,}\omega )\right\rangle =\alpha
_t(\mathbf{q},\omega )j_{f\mathbf{n}}(\mathbf{q,}\omega )/c  \label{twelve}
\end{equation}
Taking into account that $\mathbf{E}=(i\omega /c)\mathbf{A,}$ from
comparison (\ref{ten}) and (\ref{twelve}) we obtain
\begin{equation}
\alpha _t^{-1}=-\frac{i\omega }{c^2}\left( \sigma _t+\frac{pc^2}{2\pi \omega
}\right)  \label{thirteen}
\end{equation}
Using fluctuation-dissipation theorem\cite{Landau2} we get
\begin{eqnarray}
\left\langle j_{f\mathbf{n}}(\mathbf{q},\omega )j_{f\mathbf{n}^{\prime
}}^{*}(\mathbf{q}^{\prime },\omega ^{\prime })\right\rangle &=&\frac{\hbar
c^2}{2\pi i}\left( \frac 12+n(\omega )\right) \left( \alpha _t^{*-1}-\alpha
_t^{-1}\right) \delta _{\mathbf{qq}^{\prime }\,}\delta (\omega -\omega
^{\prime })  \nonumber  \label{eight} \\
\ &=&\frac{\hbar \omega }\pi \left( \frac 12+n(\omega )\right) \mathrm{Re\,}%
\sigma _t\,\delta _{\mathbf{qq}^{\prime }\,}\delta (\omega -\omega ^{\prime
}),  \label{fourteen}
\end{eqnarray}
where the Bose-Einstein factor
\[
n(\omega )=\frac 1{e^{\hbar \omega /k_BT}-1},
\]
$T$ is the temperature and \textrm{Re\thinspace }$\sigma $ is the real part
of conductivity. For $q<\left( \omega /c\right) \sqrt{\varepsilon }$ (\ref
{fourteen}) includes a term which does not depend on the conductivity, and
which result from the second term in the expression (\ref{thirteen}). This
term is associated with the black-body radiation which exist in the
dielectric even without the 2D layer. Thus this term is irrelevant to the
problem under consideration, and must be omitted. For $p-$ polarized waves
we get a similar expression as (\ref{fourteen}), except we must replace $%
\sigma _t\rightarrow \sigma _l$, where $\sigma _l$ is the longitudinal
conductivity of the layer\textbf{. }Since the $s-$ and $p-$ polarized waves
are not coupled, the average value of the product $\left\langle j_{f\mathbf{q%
}}(\mathbf{q},\omega )j_{f\mathbf{n}^{\prime }}^{*}(\mathbf{q}^{\prime
},\omega ^{\prime })\right\rangle =0$.

Let us now consider two parallel 2D electron layers separated by a distance $%
d$. We introduce two reference systems $K$ and $K^{\prime },$ with
coordinate axes $xyz$ and $x^{\prime }y^{\prime }z^{\prime }$. The $xy$- and
$x^{\prime }y^{\prime }$- planes coincide with layer \textbf{1}, with the $x$%
- and $x^{\prime }$ - axes pointing in the same direction, and the $z$- and $%
z^{\prime }$- axes pointing toward layer \textbf{2}. In the $K$ system both
layers are at rest. Assume now that in the layer \textbf{2 }the conduction%
\textbf{\ }electrons move with the drift velocity $v$, corresponding to the
current density $j_2=n_2ev,$ while no current flow in layer \textbf{1}. The $%
K^{\prime }$ reference system moves with velocity $v$ along to the $x-$ axis
relative to frame $K$. In the $K^{\prime }$ frame there is no current
density in layer \textbf{2, }while the surrounding\textbf{\ } dielectric
moves with velocity $-v\mathrm{\hat x}.$ In the $K$ frame for $z<d$ the
Maxwell equations have the form (\ref{one}) with $\mathbf{j}$ and $\mathbf{j}%
_f^{}$ replaced by $\mathbf{\QTR{mathbf}{j}_1}$ and $\mathbf{j}_{f1}^{},$
respectively. After decomposition of the components of the electromagnetic
field into a Fourier series the general solution of the Maxwell equation for
$z<d$ can be written in the form
\begin{equation}
\mathbf{E=}\left\{
\begin{array}{c}
\mathbf{v}e^{\mathrm{i}pz}+\mathbf{w}e^{-\mathrm{i}pz},\text{ }0<z<d \\
\mathbf{u}_1e^{-\mathrm{i}pz},\text{ }z<0
\end{array}
\right.  \label{fifteen}
\end{equation}
\begin{equation}
\mathbf{B=}\left\{
\begin{array}{c}
(\mathbf{\ }\left[ \mathbf{q\times v}\right] +p\left[ \mathrm{\hat z}\times
\mathbf{v}\right] )e^{\mathrm{i}pz}+(\left[ \mathbf{q\times w}\right]
-p\left[ \mathrm{\hat z}\times \mathbf{w}\right] )e^{-\mathrm{i}pz},\;0<z<d
\\
(\left[ \mathbf{q\times u}_1\right] -p\left[ \mathrm{\hat z}\times \mathbf{u}%
_1\right] )e^{-\mathrm{i}pz},\;z<0
\end{array}
\right.  \label{sixteen}
\end{equation}
where $\mathbf{v,}$ $\mathbf{w}$ and $\mathbf{u}_1$ satisfy the
transversality conditions
\begin{equation}
\mathbf{v\cdot q+}pv_z=0,\;\ \;\mathbf{w\cdot q-}pw_z=0,\text{ \ }\mathbf{u}%
_1\mathbf{\cdot q+}pu_{1z}=0  \label{seventeen}
\end{equation}
We now decompose the electromagnetic field into $s-$ and $p-$ polarized
waves. The boundary conditions at $z=0$ for $s-$ polarized waves is
determined by (\ref{nine}). For $p-$polarized waves, from (\ref{six}) one
obtain the boundary conditions
\begin{eqnarray}
E_{\mathbf{q}}(z &=&+0)=E_{\mathbf{q}}(z=-0)  \nonumber \\
\frac{\limfunc{d}E_{\mathbf{q}}}{\limfunc{d}z}\left| _{z=+0}\right. -\frac{%
\limfunc{d}E_{\mathbf{q}}}{\limfunc{d}z}\left| _{z=-0}\right. &=&-\frac{4\pi
\limfunc{i}p^2}{\varepsilon \omega }\left( \sigma _{1l}(\mathbf{q},\omega
)E_{\mathbf{q}}+j_{f1\mathbf{q}}\right)  \label{eighteen}
\end{eqnarray}

>From (\ref{nine}) and (\ref{eighteen}) we can obtain the following
equations:
\begin{equation}
v_{\mathbf{q}}+R_{1p}w_{\mathbf{q}}=-\frac{4\pi pj_{f1\mathbf{q}}}{%
\varepsilon \omega (\epsilon _{1p}+1)}  \label{nineteen}
\end{equation}
\begin{equation}
v_{\mathbf{n}}+R_{1s}w_{\mathbf{n}}=-\frac{4\pi \omega j_{f1\mathbf{n}}}{%
pc^2(\epsilon _{1s}+1)}  \label{twenty}
\end{equation}
where $v_{\mathbf{q}}=\mathrm{\hat q}\cdot \mathbf{v}$ and so on, and
\[
R_{1s(p)}=\frac{\epsilon _{1s(p)}-1}{\epsilon _{1s(p)}+1},\qquad \epsilon
_{1s}=\frac{4\pi \omega \sigma _t}{pc^2}+1,\qquad \epsilon _{1p}=\frac{4\pi
p\sigma _l}{\omega \varepsilon }+1
\]

The Maxwell equations in the $K^{\prime }-$system for $z>0$ have the same
form as (\ref{one}) with $\mathbf{j\rightarrow \QTR{mathbf}{j}}_2$ and $%
\mathbf{j}_f^{}\rightarrow \mathbf{j}_{f2}^{}$. However, to first order in $%
v/c$ the relations between \textbf{D}, \textbf{E}, and \textbf{B}, \textbf{H}
are \cite{Landau3}
\begin{equation}
\mathbf{D=}\varepsilon \mathbf{E-(}\varepsilon -1)\frac vc\mathrm{\hat x}%
\times \mathbf{B}  \label{twentyone}
\end{equation}
\begin{equation}
\mathbf{H=B-(}\varepsilon -1)\frac vc\mathrm{\hat x}\times \mathbf{E}
\label{twentytwo}
\end{equation}
After eliminating \textbf{D}, \textbf{B }and \textbf{H} from Maxwell
equations, and writing \textbf{E, j }and \textbf{j}$_f$ in Fourier series we
get
\[
\ \frac{\limfunc{d}^2\mathbf{E}^{\prime }(\mathbf{q}^{\prime })}{\limfunc{d}%
^2z}+p^{\prime 2}\mathbf{E}^{\prime }(\mathbf{q}^{\prime })-\frac{4\pi
\mathrm{i}p^{\prime 2}}{\varepsilon \omega ^{\prime }}\mathrm{\hat q}%
^{\prime }\cdot (\mathbf{j}_2\mathbf{+j}_{f2})\mathbf{e}_{\mathbf{q}^{\prime
}}\delta (z-d)-\frac{4\pi \mathrm{\hat z}}{\varepsilon \omega }\mathbf{q}%
^{\prime }\cdot (\mathbf{j}_2\mathbf{+j}_{f2})\delta ^{\prime }(z-d)
\]
\begin{eqnarray}
\ \ +\frac{4\pi \limfunc{i}\omega }{c^2}\mathbf{n}^{\prime }\mathbf{\cdot }(%
\mathbf{j}_2\mathbf{+j}_2^f)\mathbf{n}^{\prime }\delta (z-d)-\frac{4\pi
\limfunc{i}\beta q_y}{\varepsilon c}(j_{\mathbf{n}^{\prime }}\mathrm{\hat q}%
^{\prime }+j_{\mathbf{q}^{\prime }}\mathbf{n}^{\prime })\delta (z-d) &=&0
\nonumber  \label{twentythree} \\
&&  \label{twentythree}
\end{eqnarray}
where
\[
p^{\prime }=\sqrt{\left( \frac{\omega ^{\prime }}c\right) ^2\varepsilon
^{\prime }-q^{^{\prime }2},}\qquad \varepsilon ^{\prime }=\varepsilon +\frac{%
2\beta q_xc}\omega ,\qquad \beta =(\varepsilon -1)v/c
\]
Under a Lorentz transformation, with accuracy to the term linear in $v/c,$
we have $\omega ^{\prime }=\omega -q_xv$ and $\mathbf{q}^{\prime }=\mathbf{q-%
}\mathrm{\hat x}\omega v/c^2$ . Note also that $p$ is invariant under the
Lorentz transformation, i.e. $p=p^{\prime }$ . The last term in (\ref
{twentythree}) gives rise to a coupling between $s-$ and $p-$ polarized
waves. However, it can be shown \cite{Volokitin} that this coupling gives a
corrections $\sim $ $(v/c)^2$ to the frictional drag force between the
layers, so this term can be omitted. The solution of the Maxwell equations
in the $K^{\prime }$ reference frame can be written as
\begin{equation}
\mathbf{E}^{\prime }\mathbf{=}\left\{
\begin{array}{c}
\mathbf{v}^{\prime }e^{\mathrm{i}pz}+\mathbf{w}^{\prime }e^{-\mathrm{i}pz},%
\text{ \quad }0<z<d \\
\mathbf{u}_2e^{\mathrm{i}pz},\quad \quad \quad \text{ }z>d
\end{array}
\right.  \label{twentyfour}
\end{equation}
>From the boundary conditions for the $s-$ and $p-$ polarized waves, which
follow from (\ref{twentythree}), we get the equations
\begin{equation}
w_{\mathbf{q}^{\prime }}^{\prime }+R_{2p}(\mathbf{q}^{\prime },\omega
^{\prime })\mathrm{e}^{2\mathrm{i}pd}v_{\mathbf{q}^{\prime }}^{\prime }=-%
\frac{4\pi pj_{f2\mathbf{q}^{\prime }}\mathrm{e}^{\mathrm{i}pd}}{\varepsilon
\omega ^{\prime }(\epsilon _{2p}+1)}  \label{twentyfive}
\end{equation}
\begin{equation}
w_{\mathbf{n}^{\prime }}^{\prime }+R_{2s}(\mathbf{q}^{\prime },\omega
^{\prime })\mathrm{e}^{2\mathrm{i}pd}v_{\mathbf{n}^{\prime }}^{\prime }=-%
\frac{4\pi \omega ^{\prime }j_{f2\mathbf{n}^{\prime }}\mathrm{e}^{\mathrm{i}%
pd}}{pc^2(\epsilon _{2s}+1)}  \label{twentysix}
\end{equation}
The relations between the fields in the $K$ and $K^{\prime }\,$ reference
frames are determined by the Lorentz transformation. As it was shown in Ref.
\cite{Volokitin}, such a Lorentz transformation gives terms of the order $%
v/c $ which couple the $s-$ and $p-$ polarized waves but this result in a
contribution to the frictional drag of the order $(v/c)^2$. Thus we can take
this transformation in zero order in $v/c$ so that $v_{\mathbf{q}%
_{^{}}^{\prime }}^{\prime }(\omega ^{\prime })=$ $v_{\mathbf{q}}(\omega
),\;v_{\mathbf{n}^{\prime }}^{\prime }(\omega ^{\prime })=\left( \omega
^{\prime }/\omega \right) v_{\mathbf{n}}\left( \omega \right) $ and similar
equations for $\mathbf{w}$. After the transformation the solution of the
system of the equations(\ref{nineteen}, \ref{twenty}, \ref{twentyfive}, \ref
{twentysix}) take the form
\begin{equation}
v_{\mathbf{q}}=\frac{4\pi p}{\Delta _p}\left[ \frac{j_{f2\mathbf{q}^{\prime
}}\left( \mathbf{q}^{\prime },\omega ^{\prime }\right) \mathrm{e}^{\mathrm{i}%
pd}R_{1p}\left( q,\omega \right) }{\left( \epsilon _{2p}\left( \mathbf{q}%
^{\prime },\omega ^{\prime }\right) +1\right) \omega ^{\prime }}-\frac{j_{f1%
\mathbf{q}}\left( \mathbf{q,}\omega \right) }{\left( \epsilon _{1p}\left(
\mathbf{q},\omega \right) +1\right) \omega }\right]  \label{twentyseven}
\end{equation}
\begin{equation}
w_{\mathbf{q}}=\frac{4\pi p}{\Delta _p}\left[ \frac{j_{f1\mathbf{q}}\left(
q,\omega \right) \mathrm{e}^{2\mathrm{i}pd}R_{2p}\left( q^{\prime },\omega
^{\prime }\right) }{\left( \epsilon _{1p}\left( \mathbf{q},\omega \right)
+1\right) \omega }-\frac{j_{f2\mathbf{q}^{\prime }}\left( q^{\prime },\omega
^{\prime }\right) \mathrm{e}^{\mathrm{i}pd}}{\left( \epsilon _{2p}\left(
\mathbf{q}^{\prime },\omega ^{\prime }\right) +1\right) \omega ^{\prime }}%
\right]  \label{twentyeight}
\end{equation}
\begin{equation}
v_{\mathbf{n}}=\frac{4\pi \omega }{\Delta _spc^2}\left[ \frac{j_{f2\mathbf{n}%
^{\prime }}\left( q^{\prime },\omega ^{\prime }\right) \mathrm{e}^{\mathrm{i}%
pd}R_{1s}\left( q,\omega \right) }{\left( \epsilon _{2s}\left( q^{\prime
},\omega ^{\prime }\right) +1\right) }-\frac{j_{f1\mathbf{n}}\left( q,\omega
\right) }{\left( \epsilon _{1s}\left( q,\omega \right) +1\right) }\right]
\label{twentynine}
\end{equation}
\begin{equation}
w_{\mathbf{n}}=\frac{4\pi \omega }{\Delta _spc^2}\left[ \frac{j_{f1\mathbf{n}%
}\left( \mathbf{q},\omega \right) \mathrm{e}^{2\mathrm{i}pd}R_{2s}\left(
q^{\prime },\omega ^{\prime }\right) }{\left( \epsilon _{1s}(\mathbf{q}%
^{\prime },\omega )+1\right) }-\frac{j_{f2\mathbf{n}^{\prime }}\left(
\mathbf{q}^{\prime },\omega ^{\prime }\right) \mathrm{e}^{\mathrm{i}pd}}{%
\left( \epsilon _{2s}\left( \mathbf{q}^{\prime },\omega ^{\prime }\right)
+1\right) }\right]  \label{thirty}
\end{equation}
\begin{equation}
v_z=-\frac{qv_{\mathbf{q}}}p\qquad w_z=\frac{qw_{\mathbf{q}}}p
\label{thirtyone}
\end{equation}
where we have introduce the notation
\begin{eqnarray*}
\Delta _p &=&1-e^{2\mathrm{i}pd}R_{2p}\left( \mathbf{q}^{\prime },\omega
^{\prime }\right) R_{1p}\left( \mathbf{q},\omega \right) \\
\Delta _s &=&1-e^{2\mathrm{i}pd}R_{2s}\left( \mathbf{q}^{\prime },\omega
^{\prime }\right) R_{1s}\left( \mathbf{q},\omega \right)
\end{eqnarray*}

\section{Calculation of the frictional drag force between 2D systems}

The frictional drag stress $\sigma $ which acts on the conduction electrons
in layer \textbf{1 }can be obtained from the $xz-$ component of the Maxwell
stress tensor $\sigma _{ij}$, evaluated at $z=\pm 0$%
\begin{equation}
\sigma =\frac 1{8\pi }\int_{-\infty }^{+\infty }\mathrm{d}\omega \left\{
\left[ \varepsilon \langle E_zE_x^{*}\rangle +\left\langle
B_zB_x^{*}\right\rangle +c.c\right] _{z=+0}-\left[ ...\right] _{z=-0}\right\}
\label{thirtytwo}
\end{equation}
Here the $\left\langle ...\right\rangle $ denote statistical averaging over
the fluctuating current densities. The averaging is carrying out with the
aid of (\ref{fourteen}) for $s-$polarized waves and the similar equation for
$p-$ polarized waves. Note that the components of the fluctuating current
density $\mathbf{j}_{f1}$ and $\mathbf{j}_{f2}$ refer to different layers,
and are statistically independent, so that the average of their product is
zero. Expanding the electric field and magnetic induction in Fourier series
we obtain
\begin{eqnarray}
\sigma &=&\frac 1{8\pi }\int \frac{\mathrm{d}\omega \mathrm{d}^2q}{\left(
2\pi \right) ^2}\left\{ \left[ \varepsilon \langle E_z\left( \mathbf{q}%
,\omega \right) E_x^{*}\left( \mathbf{q},\omega \right) \rangle
+\left\langle B_z\left( \mathbf{q},\omega \right) B_x^{*}\left( \mathbf{q}%
,\omega \right) \right\rangle \right. \right.  \nonumber  \label{thirtythree}
\\
&&\ \ \ \left. \left. +c.c\right] _{z=+0}-\left[ ...\right] _{z=-0}\right\}
\label{thirtythree}
\end{eqnarray}
For a given value of \textbf{q }it is convenient to express the component $%
E_x$ and $B_x$ in terms of the components along the vectors \textrm{\^q} and
$\mathbf{n}$%
\begin{eqnarray}
E_x &=&(q_x/q)E_{\mathbf{q}}-(q_y/q)E_{\mathbf{n}}  \label{thirtyfour} \\
B_x &=&(q_x/q)B_{\mathbf{q}}-(q_y/q)B_{\mathbf{n}}  \label{thirtyfive}
\end{eqnarray}
After substitution of expressions (\ref{thirtyfour}-\ref{thirtyfive}) into (%
\ref{thirtythree}) and taking into account that the term which is
proportional to $q_y$ is equal to zero \cite{Volokitin}, we obtain
\begin{eqnarray}
\sigma &=&\frac 1{8\pi }\int \frac{\mathrm{d}\omega \mathrm{d}^2q}{\left(
2\pi \right) ^2}\frac{q_x}q\left\{ \left[ \varepsilon \langle E_z\left(
\mathbf{q},\omega \right) E_{\mathbf{q}}^{*}\left( \mathbf{q},\omega \right)
\rangle +\left\langle B_z\left( \mathbf{q},\omega \right) B_{\mathbf{q}%
}^{*}\left( \mathbf{q},\omega \right) \right\rangle \right. \right.
\nonumber \\
&&\ \ \ \left. \left. +c.c\right] _{z=+0}-\left[ ...\right] _{z=-0}\right\}
\label{thirtysix}
\end{eqnarray}
where
\begin{equation}
E_z(z=+0)=(v_z+w_z)=(q/p)(w_{\mathbf{q}}-v_{\mathbf{q}})=(qp^{*}/\mid p\mid
^2)(w_{\mathbf{q}}-v_{\mathbf{q}})  \label{thirtyseven}
\end{equation}
\begin{equation}
E_z(z=-0)=u_{1z}=(q/p)u_{\mathbf{q}}=(q/p)(w_{\mathbf{q}}+v_{\mathbf{q}})
\label{thirtyeight}
\end{equation}
\begin{equation}
E_{\mathbf{q}}(z=+0)=E_{\mathbf{q}}(z=-0)=v_{\mathbf{q}}+w_{\mathbf{q}}
\label{thirtynine}
\end{equation}
\begin{equation}
B_z(z=+0)=(qc/\omega )(v_{\mathbf{n}}+w_{\mathbf{n}})=B_z(z=-0)=(qc/\omega
)u_{1\mathbf{n}}  \label{fourty}
\end{equation}
\begin{equation}
B_{\mathbf{q}}(z=+0)=(pc/\omega )(w_{\mathbf{n}}-v_{\mathbf{n}})
\label{fourtyone}
\end{equation}
\begin{equation}
B_{\mathbf{q}}(z=-0)=(pc/\omega )u_{1\mathbf{n}}  \label{fourtytwo}
\end{equation}
After substituting these expressions into formula (\ref{thirtysix}) we
obtain
\begin{eqnarray}
\sigma &=&\frac 1{16\pi ^3}\int_0^{+\infty }\mathrm{d}\omega \int \mathrm{d}%
^2qq_x\left( \frac \varepsilon {\mid p\mid ^2}\left[ (p+p^{*})(\langle \mid
w_{\mathbf{q}}\mid ^2\rangle -\langle \mid v_{\mathbf{q}}\mid ^2\rangle
\right. \right.  \nonumber \\
&&\ \ \ \left. -\langle \mid v_{\mathbf{q}}+w_{\mathbf{q}}\mid ^2\rangle
)+(p-p^{*})\langle (v_{\mathbf{q}}w_{\mathbf{q}}^{*}-v_{\mathbf{q}}w_{%
\mathbf{q}}^{*})\rangle \right]  \nonumber \\
&&\ \ \ +\left( \frac c\omega \right) ^2\left[ (p+p^{*})(\langle \mid w_{%
\mathbf{n}}\mid ^2\rangle -\langle \mid v_{\mathbf{n}}\mid ^2\rangle
-\langle \mid v_{\mathbf{n}}+w_{\mathbf{n}}\mid ^2\rangle )\right.  \nonumber
\label{fourtythree} \\
&&\ \ \ \left. \left. -(p-p^{*})\langle (v_{\mathbf{n}}w_{\mathbf{n}}^{*}-v_{%
\mathbf{n}}w_{\mathbf{n}}^{*})\rangle \right] \right)  \label{fourtythree}
\end{eqnarray}
where we integrate only over positive values of $\omega $, which gives an
extra factor of two.

Substituting $($\ref{twentyseven}) and (\ref{thirtyone}) into (\ref
{fourtythree}) and taking into account that $p=p^{*}$ for $q<\omega /c$ and $%
p=-p^{*}$ for $q>\omega /c,$ we obtain
\begin{eqnarray}
\sigma &=&\frac \hbar {8\pi ^3}\int_0^\infty \mathrm{d}\omega
\int_{q<(\omega /c)\sqrt{\varepsilon }}\mathrm{d}^2qq_x  \nonumber \\
&&\times  \left[ \frac{T_{1p}(\omega )T_{2p}(\omega -q_xv)(n(\omega
-q_xv)-n(\omega ))}{\mid 1-\mathrm{e}^{2\mathrm{i}pd}R_{1p}(\omega
)R_{2p}(\omega -q_xv)\mid ^2}  \nonumber \right. \\
&&\left. -\frac{T_{1p}(\omega )(\mid 1-R_{1p}(\omega )\mid ^2)+(\mid 1-%
\mathrm{e}^{\mathrm{i}pd}R_{1p}(\omega )\mid ^2)(n(\omega )+1/2)}{\mid 1-%
\mathrm{e}^{2\mathrm{i}pd}R_{1p}(\omega )R_{2p}(\omega -q_xv)\mid ^2}\right]
\nonumber \\
&&\ \ \ \left.  +\frac \hbar {2\pi
^3}\int_0^\infty \mathrm{d}\omega \int_{q>(\omega /c)\sqrt{\varepsilon }}%
\mathrm{d}^2qq_x\mathrm{e}^{-2\mid p\mid d}  \nonumber \right. \\
&&\times  \frac{\mathrm{Im}R_{1p}(\omega )\mathrm{Im}%
R_{2p}(\omega -q_xv)}{\mid 1-\mathrm{e}^{-2\mid p\mid d}R_{1p}(\omega
)R_{2p}(\omega -q_xv)\mid ^2}(n(\omega -q_xv)-n(\omega ))
\nonumber  \label{fourtyfour} \\
&&\left. +\left[ p\rightarrow s\right]  \label{fourtyfour} \right.
\end{eqnarray}
where
\begin{eqnarray*}
T_{ip}(\omega ) &=&1-\mid R_{ip}\mid ^2-\mid 1-R_{ip}\mid ^2=\frac{16\pi
\mathrm{Re}\sigma _l(\omega )p}{\omega \varepsilon |\epsilon _{il}+1|^2} \\
T_{is}(\omega ) &=&1-\mid R_{is}\mid ^2-\mid 1-R_{is}\mid ^2=\frac{16\pi
\mathrm{Re}\sigma _t(\omega )\omega }{pc^2|\epsilon _{it}+1|^2}
\end{eqnarray*}
The first integral in $($\ref{fourtyfour}) is the contribution to the
frictional drag force from propagating electromagnetic waves. This integral
contains terms which formally diverge upon integration over $\omega .$ These
terms are proportion to $\omega ^{-1}$ at large frequencies and appear as a
result of the expansion of the reflection factor $R_{2p(s)}(\omega -q_xv)$
in power series in $q_xv,$ and upon performing the $q$ integration. A
similar divergence also occur in the derivation of the static van der Waals
interaction \cite{Lifshitz}, and result from zero point vacuum fluctuations
of electromagnetic field. The solution of this problem consists of
subtraction from the integrand the terms which do not depend from
separation between layers $d$ in the limit $d$ $\rightarrow \infty .$ In our
case this procedure consists of subtraction from the first integrand in (%
\ref{fourtyfour}) the same expression taken at $T=0\,K$ \ and with
denominator equal to unity. The second term in (\ref{fourtyfour}) is derived
from the evanescent field.

\section{Some limiting cases}

Consider distances $d<<d_W\sim c\hbar /k_BT$ (at $T=3\;K$ \ we have\ \ $d_W$
$\sim 10^6\,\mathrm{\AA }$). In this case we can neglect by the first
integral in (\ref{fourtyfour}), put $p\approx \mathrm{i}q$ and extend the
integral over $\mathbf{q}$ to the whole $q-$ plane. Using these
approximations, the second integral in (\ref{fourtyfour}) can be written as
\cite{Volokitin}
\begin{eqnarray}
\sigma &=&\frac \hbar {2\pi ^3}\int_{-\infty }^{+\infty }\mathrm{d}%
q_y\int_0^\infty \mathrm{d}q_xq_x\mathrm{e}^{-2qd}  \nonumber \\
&&\ \times \left\{ \int_0^\infty \mathrm{d}\omega [n(\omega )-n(\omega
+q_xv)]\right.  \nonumber \\
&&\ \times \left[ \left( \frac{\mathrm{Im}R_{1p}(\omega )\mathrm{Im}%
R_{2p}(\omega +q_xv)}{\mid 1-\mathrm{e}^{-2qd}R_{1p}(\omega )R_{2p}(\omega
+q_xv)\mid ^2}+(1\leftrightarrow 2)\right) +(s\rightarrow p)\right] 
\nonumber \\
&&\ -\int_0^{q_xv}\mathrm{d}\omega [n(\omega )+1/2]\left[ \left( \frac{%
\mathrm{Im}R_{1p}(\omega -q_xv)\mathrm{Im}R_{2p}(\omega )}{\mid 1-\mathrm{e}%
^{-2qd}R_{1p}(\omega -q_xv)R_{2p}(\omega )\mid ^2}+(1\leftrightarrow
2)\right) \right.  \nonumber \\
&&\left. \left. +(s\rightarrow p)\right] \right\}  \label{fourtyseven}
\end{eqnarray}
The second term in this expression is proportional to $v^2$ as$%
\,v\rightarrow 0$ and can be neglected in the limit of small $v.$ In the
first term we can use approximation 
\[
n(\omega )-n(\omega +q_xv)\approx -q_xv\frac{\mathrm{d}n}{\mathrm{d}\omega }=%
\frac{\mathrm{e}^{\hbar \omega /k_BT}}{(\mathrm{e}^{\hbar \omega /k_BT}-1)^2}%
\frac{\hbar q_xv}{k_BT} 
\]
Thus 
\begin{eqnarray}
\sigma &=&\frac{\hbar v}{2\pi ^2}\int_0^\infty \mathrm{d}qq_{}^3\mathrm{e}%
^{-2qd}\int_0^\infty \mathrm{d}\omega \left( -\frac{\mathrm{d}n}{\mathrm{d}%
\omega }\right) \left\{ \frac{\mathrm{Im}R_{1p}(\omega )\mathrm{Im}%
R_{2p}(\omega )}{\mid 1-\mathrm{e}^{-2qd}R_{1p}(\omega )R_{2p}(\omega )\mid
^2}+[p\rightarrow s]\right\}  \nonumber \\
&&  \label{fourtyeight}
\end{eqnarray}
Note that $\sigma $ is linear in the velocity $v$.
Let us describe the 2D layers in RPA approximation. Thus for $q<k_F$
(corresponding to separations $d> k_F^{-1},$where $k_F$ is the Fermi wave
vector of the degenerate electron gas system; for 2D electron layer with
electron density $n_s\approx 1.5\cdot 10^{11}\mathrm{cm}^{-2},\,\,k_F=(2\pi
n_s)^{1/2}\sim 10^6\mathrm{cm}^{-1}$) the transverse and longitudinal parts
of the conductivity for 2D electron layer can be written in the form \cite
{Stern,Persson}
\begin{equation}
\sigma _l=\frac{\mathrm{i}\omega e^2n_s}{q^2\epsilon _F}\left\{ \frac{\omega
\overline{u}}{(\omega +\mathrm{i}\gamma )\sqrt{\overline{u}^2-1}-\mathrm{i}%
\gamma \overline{u}}-1\right\}  \label{fourtynine}
\end{equation}
\begin{equation}
\sigma _t=-\frac{2\mathrm{i}e^2n_s\overline{\,u}(\sqrt{\overline{\,u}^2-1}-%
\overline{\,u})}{m^{*}(\omega +\mathrm{i}\gamma )}  \label{fifty}
\end{equation}
where $\overline{\,u}=(\omega +\mathrm{i}\gamma )/qv_F,\,\gamma =1/\tau
,\,v_F=\hbar k_F/m^{*}$ \thinspace is the Fermi velocity, $\tau $ is a
relaxation time, $\epsilon _F=\hbar ^2k_F^2/2m^{*}$ is the Fermi energy. In
the experiment \cite{Gramila1,Gramila2} $m^{*}=0.067m_e,v_F=1.6\times 10^7%
\mathrm{cm/s,\,}\epsilon _F\sim 60\,\mathrm{K}$ and the mobility $\mu \sim $
2$\times 10^6\mathrm{cm}^2/\mathrm{Vs,}$ so that $\tau \sim 7.6\times
10^{-11}\mathrm{s.}$ \thinspace \thinspace Let us divide the integration
over $0<q<\infty $ into the two parts $0<q<\omega /v_F$ and $\omega
/v_F<q<\infty .$ In the first part of integration $\overline{u}>1,$ and
taking the limit $\overline{u}>>1$ we obtain in this limit the Drude formula
for conductivity
\begin{equation}
\sigma _l=\sigma _t=\frac{\mathrm{i}e^2n_s}{m^{*}(\omega +\mathrm{i}\gamma )}
\label{fiftyone}
\end{equation}
In the second part of integration $\overline{u}<1$ and taking the limit $%
\overline{u}<<1$ we obtain
\begin{equation}
\sigma _l=\frac{\omega e^2n_s}{q^2\epsilon _F}(u-\mathrm{i})
\label{fiftytwo}
\end{equation}
\[
\sigma _t=\frac{e^2n_sv_F\,}{\epsilon _Fq}\quad \quad ;
\]
where we put $\gamma $ equal to zero because it gives only a small
contribution in this limit.

Let us consider the case of small separation $d$ when $a=(2k_BTd/\hbar
v_F)<1 $. Introducing the dimensionless variables $q=x/2d$ and$\;\omega
=(k_BT/\hbar )y$ we obtain in this limit for $ay<x<\infty $
\begin{equation}
R_p=\frac{\lambda _p(x+\mathrm{i}ay)}{x^2+\lambda _p(x+\mathrm{i}ay)}\approx
1-\frac{x^2}{\lambda _p(x+\mathrm{i}ay)}  \label{fiftythree}
\end{equation}
\begin{equation}
R_s=\frac{\mathrm{i}\lambda _sy}{\mathrm{i}\lambda _sy-x^2}
\label{fiftyfour}
\end{equation}
and for $0<x<ay$%
\begin{equation}
R_p=\frac{\lambda _p^{\prime }x}{\lambda _p^{\prime }x-2y^2-2\mathrm{i}%
y\delta }  \label{fiftyfive}
\end{equation}
\begin{equation}
R_s=\frac{\lambda _s^{\prime }y}{2xy+\lambda _s^{\prime }y+\mathrm{i}%
2x\delta }  \label{fiftysix}
\end{equation}
where
\[
\lambda _p=\frac{8\pi e^2n_sd}{\varepsilon m^{*}v_F^2},\quad \lambda _s=8\pi
ad\left( \frac{e^2n_s}{m^{*}c^2}\right) ,\,\,\,\,\,\,\lambda _p^{\prime }=%
\frac{2\pi n_se^2}{\varepsilon m^{*}d}\left( \frac \hbar {k_BT}\right)
^2,\quad \lambda _s^{\prime }=\frac{8\pi n_se^2d}{m^{*}c^2}
\]
We note that the expression (\ref{fiftyfive}) has pole at
\begin{equation}
\omega ^2=\frac{2\pi n_se^2}{\varepsilon m^{*}}q  \label{fiftyseven}
\end{equation}
what corresponds to the plasmon excitations \cite{Ando}. After substituting (%
\ref{fiftythree}- \ref{fiftysix}) in (\ref{fourtyeight}) we obtain for the
frictional drag rate
\begin{equation}
\tau _{Dp}^{-1}\approx 0.2360\frac{(kT)^2}{\hbar \epsilon
_F(q_{TF}d)^2(k_Fd)^2}+10\left( \frac{k_BT}{\epsilon _F}\right) ^5\left(
\frac{k_BT}{\epsilon _{TF}}\right) ^2\gamma  \label{fiftyeight}
\end{equation}
\begin{equation}
\tau _{Ds}^{-1}\approx 3.3\cdot 10^{-5}\left( \frac{k_BT}{m^{*}c^2}\right)
\left( 4\frac{k_BT}\hbar +\gamma \right)  \label{fiftynine}
\end{equation}
where $\tau _{Dp}^{-1}$ and $\tau _{Ds}^{-1}$ are the contributions from $s-$
and $p-$ polarized waves, respectively, $q_{TF}=2e^2m^{*}/\hbar
^2\varepsilon $ is the single-layer Thomas-Fermi screening wavevector, $%
\epsilon _{TF}=\hbar ^2q_{TF}^2/2m^{*}.$The first term in (\ref{fiftyeight})
agrees with the result of Gramila \textit{at al} \cite{Gramila2} and Persson
and Zhang \cite{Persson and Zhang}. From comparison (\ref{fiftyeight}) and (%
\ref{fiftynine}) it follows that for
\begin{equation}
n_s<n_c\sim 10^2\left( \frac{mk_BT}{\pi \hbar ^2}\right) \left( \frac{%
\varepsilon ^4\hbar ^2k_BT}{m^{*}e^4}\right) ^{1/5}  \label{sixty}
\end{equation}
the contribution from $p-$ polarized waves exceeds the contribution from $s-$
polarized waves for all distances $d<\hbar v_F/k_BT$. However for $n_s>n_c$
the contribution from $s-$ polarized waves will dominate for $%
d>10(\varepsilon /n_s)^{1/2}$. For example, for $T=3\,K$ and for the
conditions of experiment of Ref.\cite{Gramila1,Gramila2} $n_c\sim
10^{12}\,cm^{-2}$ and we find that in this case the retardation effects are
small. However retardation effects are important for high electron
densities. For example, assuming that $\varepsilon = 1$ and
$n_s \approx 10^{15} {\rm cm}^{-2}$, which correspond to
about 1 monolayer of silver,
we find that the contribution to frictional drag from the retardation
effects will dominate for $d>15\ {\rm \AA} $. This is illustrated in Fig. 2a
which shows the
shear stress when the relative velocity
$v= 1 {\rm m/s}$. We have
performed calculations for two different temperatures, $T=273 {\rm K}$ and $77
{\rm K}$,
and the $s$ and $p$-wave contributions are shown separately. In Fig. 2b we show
the same quantity for two quantum wells at $T= 3 {\rm K}$ and with $n_s = 1.5
\times 10^{11}
{\rm cm}^{-2}$, $m^* = 0.067 m_e$, $v_F= 1.6\times 10^7 {\rm  cm/s}$ and $\tau =
7.6 \times 10^{-11} {\rm s}$, and with $\varepsilon = 1$. In this case the
$p$-wave contribution dominates for $d < 1000 \ {\rm \AA}$.

Let us estimate the voltage $U_1$ induced in a thin silver film (layer 1) (open
circuit) when a
current flow in another parallel silver film (layer 2).
A voltage difference of order $1 \ {\rm pV}$ can be measured with standard
equipment so that if $U_1$ is of order ${\rm pV}$ or larger, it is possible
to probe retardation effects with this experimental setup. If $L$ denote the
length
of the metallic films (assumed identical)
in the direction of the driving current, then $U_1 =L E_1$ and
$U_2 = LE_2 = L J_2/\sigma_2$ where
$\sigma_2 = n_2 e^2 \tau_2/m^*$ is the conductivity ($\tau_2$ is a Drude
relaxation time and
$m^*$ the electron effective mass). Thus, using the equation (see introduction)
$\gamma = n_1 n_2 e^2 E_1/J_2$ with $E_1/J_2 = U_1/(\sigma_2 U_2) = (U_1/U_2)
m^*/(n_2 e^2 \tau_2)$
gives $U_1 = (\gamma \tau /m^* n_1) U_2$. In a typical case $\tau = 4\times
10^{-14} \ {\rm s}$
and $n_1 \approx 10^{15} \ {\rm cm}^{-2}$, and from Fig. 2a, $\gamma \approx
10^{-6} \ {\rm
N s /m^2}$, giving $U_1 \approx 10^{-8} \ U_2$. Thus if the applied voltage $U_2
\approx 1 \ {\rm V}$,
the induced voltage would be of order $10 \ {\rm pV}$, which should be possible
to measure.

\section{Frictional drag between 3D systems}

For high electron densities, when the thickness of the layers $%
h>>n^{-1/3},$ where $n$ is a volume electron density, the electrons behave
as in 3D systems. It was shown in Ref.\cite{Volokitin} that
for 3D systems the frictional drag stress is also given by formula (\ref
{fourtyeight}), where the electromagnetic reflection coefficients
\begin{equation}
R_{ip}=\frac{\varepsilon _ip-\varepsilon s_i}{\varepsilon _ip+\varepsilon s_i%
},\,\,\;\;\;\;\;R_{is}=\frac{p-s_i}{p+s_i},  \label{sixtyone}
\end{equation}
where $\varepsilon _i\,$ is the complex dielectric constant for layer
\textbf{i,}
\begin{equation}
s_i=\sqrt{\frac{\omega ^2}{c^2}\varepsilon _i-q^2}  \label{sixtytwo}
\end{equation}

Consider two identical 3D layers described by the dielectric function
\begin{equation}
\varepsilon (\omega )=1-\frac{\omega _p^2}{\omega (\omega +\mathrm{i}/\tau )}%
,  \label{sixtythree}
\end{equation}
where $\tau $ is the Drude relaxation time and $\omega _p$ the plasma
frequency.
For $k_BT>\hbar \gamma ,$ for small frequencies and $d<c/\omega _p(\hbar
\gamma /k_BT)^{1/2}$%
\begin{eqnarray*}
\mathrm{Im}R_p &\approx &\frac{2\varepsilon \omega }{\omega _p^2\tau }%
,\;\;\;\;\;\;\;\;\;\;\;\;\;\;\mathrm{Re}R_p\approx 1,\;\;\;\; \\
\mathrm{ReR}_s &\approx &0,\;\;\mathrm{Im}R_s\approx 4\left( \frac{\omega _p}%
c\right) ^2\frac \omega {\gamma q^2},\;\;{\rm for}\ \;q^2>\left( \frac{\omega
_p}%
c\right) ^2\frac \omega \gamma ,
\end{eqnarray*}
then, taking into account that for 3D systems $n_s=nh,$ (\ref
{fourtyeight}) gives
\begin{equation}
\tau _{Dp}^{-1}=13.32\frac{(\varepsilon k_BT)^2}{\hbar \epsilon
_F(k_Fd)^2(k_{TF}d)^2(k_Fh)(\omega _p\tau )^2}  \label{sixtyfour}
\end{equation}
\begin{equation}
\tau _{Ds}^{-1}=\frac{e^2(k_BT)^2(\omega _p\tau )^2}{8\pi \hbar h(m^{*}c^2)^2%
},  \label{sixtyfive}
\end{equation}
where $k_{TF}^2=6\pi ne^2/\epsilon _F$ is the 3D Thomas-Fermi screening
wavevector, and $k_F=(3\pi ^2n)^{1/3}$ is the 3D Fermi wavevector. From
comparison (\ref{sixtyfour}) and (\ref{sixtyfive}) it follows that,
for $d>\varepsilon ^{1/2}(c/\omega _p)(\omega _p\tau )^{-1}$,
the $s-$wave contribution exceeds the $p-$wave contribution
Thus, in the case of the dissipative van der Waals interaction between 3D
bodies, retardation effects become important for much shorter distances
in comparison with conservative one, when the retardation effects become
important for $d>c/\omega _p$ \cite{Lifshitz}.

Fig. 3 shows the calculated shear stress for two semi-infinite silver bodies
moving with the relative velocity $v= 1{\rm m/s}$ parallel to the flat
surfaces. Results are shown for the $s$- and $p$-wave contribution, where in
the latter case
we have taken into account non-local effects [the dashed lines shows the
result when the local (long-wavelength) dielectric function is used]. Results
are shown
for two different temperatures, $T= 70 {\rm K}$ and $300 {\rm K}$, and the
observed temperature
dependence reflect both that of the temperature prefactor $T^2$ in the
expression for the shear stress, as well as the temperature dependence of the

\section{Relation between friction and heat transfer}

The frictional shear stress studied above is closely related to the heat
transfer
from one solid to another when the solids have different temperatures. For
large
separation the heat transfer is given by the Stefan's law
\begin{equation}
J_z = {\pi^2 k_B^4 \over 60 \hbar^3 c_0^2} \left (T_1^4 -T_2^4\right ) \label{seventy}
\end{equation}
where $T_1$ and $T_2$ are the temperatures of solid 1 and 2, respectively.
This formula correspond to emission of real photons. However, for short
separation
$d$ it is possible for the evanescent near field to transfer energy from one
solid to
the other. This correspond to photon-tunneling. In general, the heat
flux (energy flow per unit area and unit time) is given by a formula very
similar to that
for the frictional stress \cite{Van Hove,Volokitin1} :

\begin{eqnarray}
J_z &=&\frac \hbar {8\pi ^3}\int_0^\infty \mathrm{d}\omega \omega
\int_{q<\omega /c }\mathrm{d}^2q  \nonumber \\
&&\times\left[
\frac{(1-\mid R_{1p}(\omega )\mid ^2)(1-\mid R_{2p}(\omega)\mid ^2)%
(n_{1}(\omega )-n_{2}(\omega)}{\mid 1-%
\mathrm{e}^{2\mathrm{i}pd}R_{1p}(\omega )R_{2p}(\omega )\mid ^2}\right]
\nonumber \\
&&  +\frac \hbar {2\pi
^3}\int_0^\infty \mathrm{d}\omega \omega\int_{q>\omega /c}%
\mathrm{d}^2q\mathrm{e}^{-2\mid p\mid d}  \nonumber \\
&&\times  \frac{\mathrm{Im}R_{1p}(\omega )\mathrm{Im}%
R_{2p}(\omega )}{\mid 1-\mathrm{e}^{-2\mid p\mid d}R_{1p}(\omega
)R_{2p}(\omega)\mid ^2}(n_{1}(\omega)-n_{2}(\omega ))
\nonumber  \label{fourtyfour} \\
&&\left. +\left[ p\rightarrow s\right]  \label{seventyone} \right.
\end{eqnarray}
where
\begin{equation}
n_1(\omega) = \left (e^{\hbar \omega /k_BT_1}-1\right )^{-1}
\end{equation}
is the Bose-Einstein factor of solid 1 and similar for $n_2$.
Fig. 4a shows the heat transfer between two semi-infinite silver bodies
separated
by the distance $d$ and at the temperatures
$T_1 = 273 {\rm K}$ and $T_2 = 0 {\rm K}$. The $s$ and $p$-wave contribution
are shown separately,
and the $p$-wave contribution has been calculated using non-local optics (the
dashed
line shows the result using local optics). It is remarkable how important the
$s$-contribution is even for short distances.
The detailed distance dependence of $J_z$ has been studied
by Ploder and Van Hove within the local optics approximation, and will not
be repeated here. The nonlocal optics contribution to $(J_z)_p$,
which is important only for $d < l$ (where $l$ is the electron mean free path
in the bulk),
is easy to calculate for free electron like metals.
The non-local contribution to ${\rm Im} R_p$ is given by \cite{Persson and Zhang}
$$\left ( {\rm Im} g \right )_{\rm surf} = 2 \xi {\omega\over \omega_p}
{q\over k_F}$$
Using this expression for ${\rm Im} R_p$ in (\ref{seventyone} ) gives the (surface)
contribution:
$$J_{\rm surf} \approx {\hbar \xi^2 \over \omega_p^2 k_F^2 d^4}
\left ({k_BT_1 \over \hbar}\right )^4 f(T_1/T_2)$$
where
$$f(T_1/T_2) = {1\over 4\pi^2} \int_0^\infty dx {x^3e^{-x} \over \left
(1-e^{-x}\right )^2}
\int_0^\infty dy y^3 \left ({1\over 1-e^{-y}}
-{1\over 1-e^{-(T_1/T_2)y}}\right )$$
$$= 0.1827
\int_0^\infty dy y^3 \left ({1\over 1-e^{-y}}
-{1\over 1-e^{-(T_1/T_2)y}}\right )\rightarrow 1.186$$
as $T_2/T_1 \rightarrow 0$. Note from Fig. 4a that the local optics
contribution to
$(J_z)_p$ depends nearly linearly on $1/d$ in the studied distance interval,
and that this
contribution is much smaller than the $s$-wave contribution. Both these
observations
differ from Ref. \cite{Pendry}, where it is stated that the $s$ contribution can be
neglected for
small distances and that the $p$-wave contribution (within local optics) is
proportional to
$1/d^2$ for small distances. However, for the very high-resistivity
materials, the $p$-wave contribution becomes much more important,
and a crossover to a $1/d^2$-dependence of $(J_z)_p$ is observed at very short
separations
$d$. This is illustrated in Fig. 4b and 4c, which have been calculated with the
same
parameters as in Fig. 4a, except that the electron mean free path has been
reduced from
$l= 560 \ {\rm \AA}$ (the electron mean free path for silver at room
temperature)
to $20 \ {\rm \AA}$ (roughly the electron mean free path in lead at room
temperature)
(Fig. 4b) and $3.4 \ {\rm \AA}$
(of order the lattice constant, representing the minimal possible mean free
path)
(Fig. 4c). Note that when $l$ decreases, the $p$ contribution to the heat
transfer increases while
the $s$ contribution decreases. Since the mean free path cannot be much smaller
than the lattice constant, the result in Fig. 4c
represent the largest possible $p$-wave
contribution for normal metals. However, the $p$-wave contribution may be even
larger for
other materials, e.g., semimetals, with lower carrier concentration than in
normal metals.
This fact has already been pointed out by Pendry: the p-wave contribution for
short distances
is expected to be maximal when
the function
$${\rm Im} R_p \approx {\rm Im} {\epsilon -1 \over \epsilon +1}
={\rm Im} \left [ 1-2{\omega \over \omega_p}\left ({\omega \over
\omega_p}+{i\over \omega_p \tau}
\right )\right ]^{-1}$$
is maximal with respect to variations in $1/\tau$. This gives:
$$\omega_p \tau = {2k_BT \over \hbar \omega_p}$$
where we have used that typical frequencies $\omega \sim k_B T /\hbar$. Since
the
DC resistivity $\rho = 4\pi/(\omega_p^2\tau)$ we get (at room temperature)
$\rho \approx 2 \pi \hbar /k_BT
\approx 0.14 \ \Omega {\rm cm}$.

\section{Summary and conclusion}

We have used a general theory of a fluctuating electromagnetic field to
calculate the frictional drag force between 2D and 3D
electron systems. The separation $d$ between the parallel electron layers is
assumed to be so large that the only interaction between the layers is via
the electromagnetic field associated with \textit{thermal }and \textit{%
quantum} fluctuations in the layers; the resulting friction force
can be considered as the dissipative part of the
van der Waals interaction. A general formula has been obtained, in which the
frictional drag force is expressed through the electromagnetic reflection
coefficients for $s$ and $\;p-$waves. We have found that the
non-retarded Coulomb interaction, connected with evanescent $p-$polarized
waves, is the predominant process for small layer separations and small
electron densities. For high electron densities
retardation effects (connected with evanescent $s-$polarized waves)
become very important, and we have suggested a new experiment,
involving thin metallic films, where the theory can be tested.
We have shown that retardation effects are even more important
for interaction between 3D electron systems.
For very large separations the interaction is dominated by the
traveling electromagnetic waves, which results from black-body radiation.
However, the latter interaction appears negligible in comparison with phonon
mediated process. Finally, we have pointed out the close relation between
heat transfer and friction.
\vskip 0.5cm
\textbf{Acknowledgment }

A.I.V acknowledges financial support from DFG . B.N.J. P acknowledges
financial support from BMBF.
\vskip 1cm

FIGURE CAPTIONS

Fig. 1 Left: a metallic block sliding relative to the metallic substrate with
the
velocity $v$. An electronic frictional shear stress $\sigma$ will act on the
block (and on the
substrate). Right: The shear stress $\sigma$ can be measured if instead of
sliding the
upper block, a voltage $U_2$ is applied to the block resulting in a drift
motion of the
conduction electrons (velocity $v$). The resulting frictional stress on the
substrate
electrons will generate a voltage difference $U_1$ (proportional to $\sigma$)
as indicated in the figure, which
can be measured experimentally.

Fig. 2 The shear stress as a function of the
distance $d$ between the surfaces (the log-function is with basis 10). (a) For
$\sim$monolayer-films of silver for two different temperatures. The $s$ and
$p$-wave contributions are shown separate. In the calculation $\tau = 4\times
10^{-14}
{\rm s}$ and $20\times 10^{-14} {\rm s}$ for $T=273 {\rm K}$ and 77 K,
respectively.
We have assumed $n_s=1.05\times 10^{19} {\rm m^{-2}}$, $m^* = m_e$, and
$v_F = 1.4 \times 10^6 \ {\rm m/s}$. (b) For quantum wells at $T= 3 \ {\rm K}$.
In the calculation $\tau = 7.6\times 10^{-11} \ {\rm s}$,
$n_s = 1.5 \times 10^{15} \ {\rm m}^{-2}$,
$m^*= 0.067 m_e$, and $v_F = 1.6 \times 10^5 \ {\rm m/s}$.

Fig. 3 The shear stress as a function of the distance $d$ between the surfaces
of
two semi-infinite silver bodies. The $s$ and $p$ wave contributions are
shown separately and for two different temperatures,
$T=70 \ {\rm K}$ and $300 \ {\rm K}$.
The $p$-wave contribution has been calculated both
using a local dielectric function (dashed lines) and using a theory which takes
into account
nonlocality within the jellium model.

Fig. 4 (a) The heat transfer flux between two semi-infinite silver bodies, one
at
temperature $T_1 = 273 \ {\rm K}$ and another at $T_2 = 0 \ {\rm K}$. (b) The
same as
(a) except that we have reduced the electron relaxation time $\tau$ for solid 1
from
a value corresponding to a mean free path $v_F \tau = l = 560 \ {\rm \AA}$ to
$20 \ {\rm \AA}$. (c) The same as (a) except that we have reduced $l$ to $3.4 \
{\rm \AA}$.

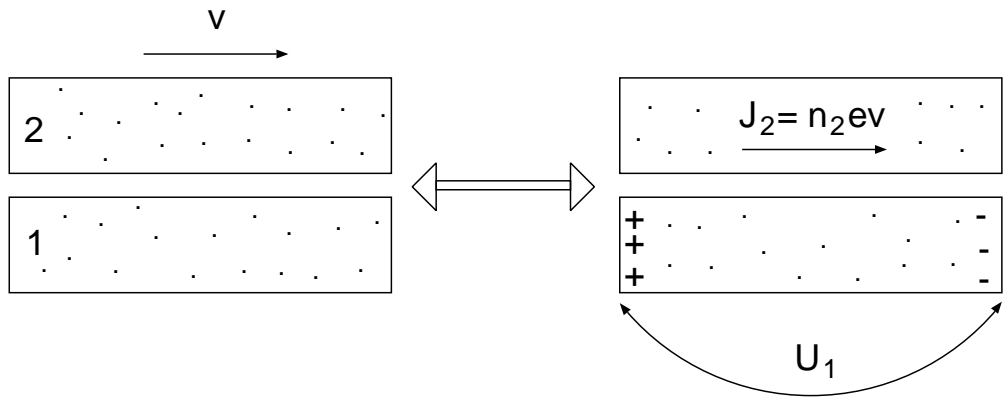
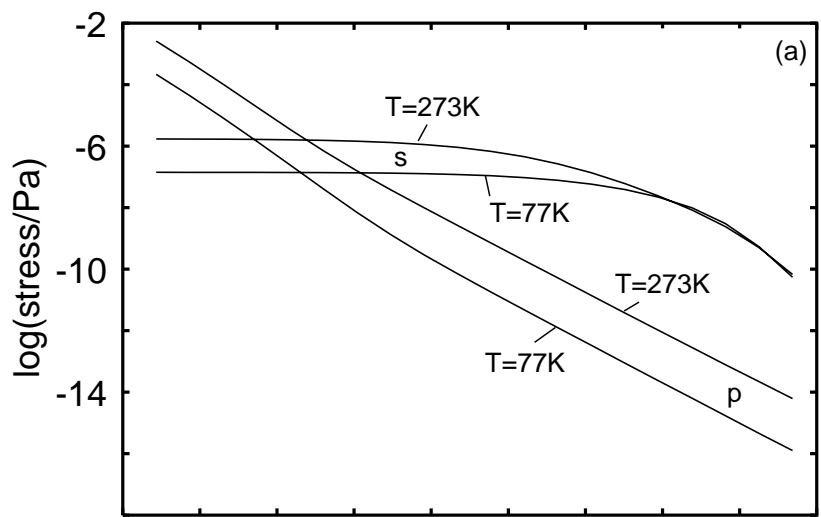
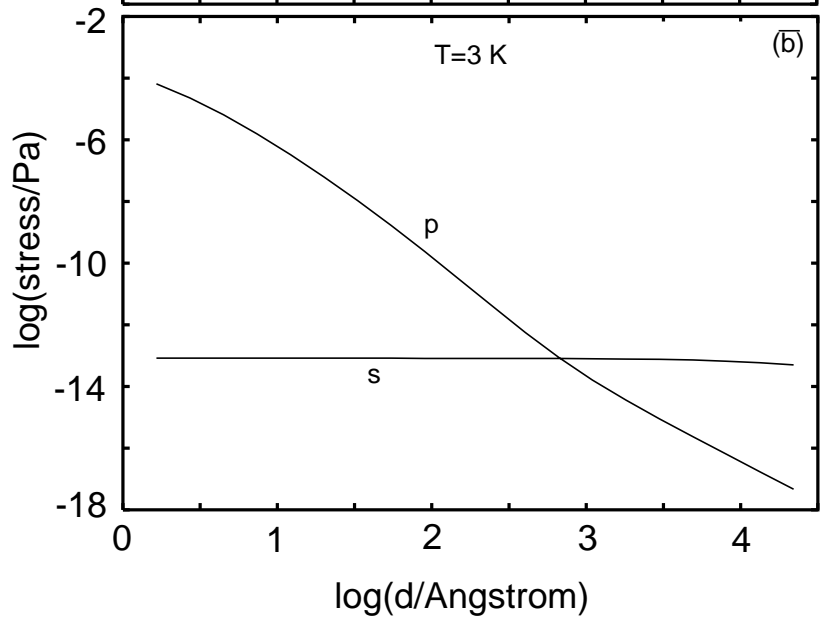
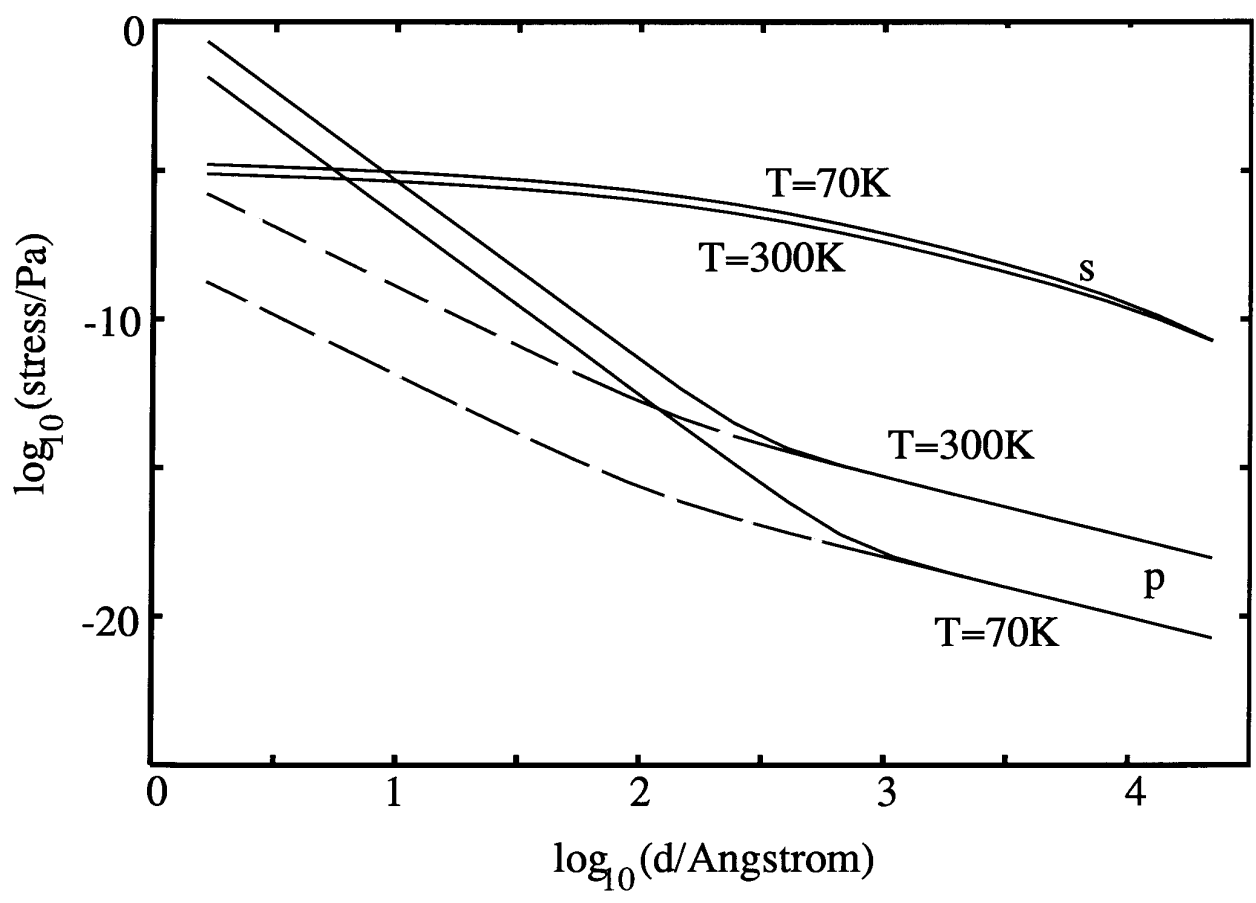
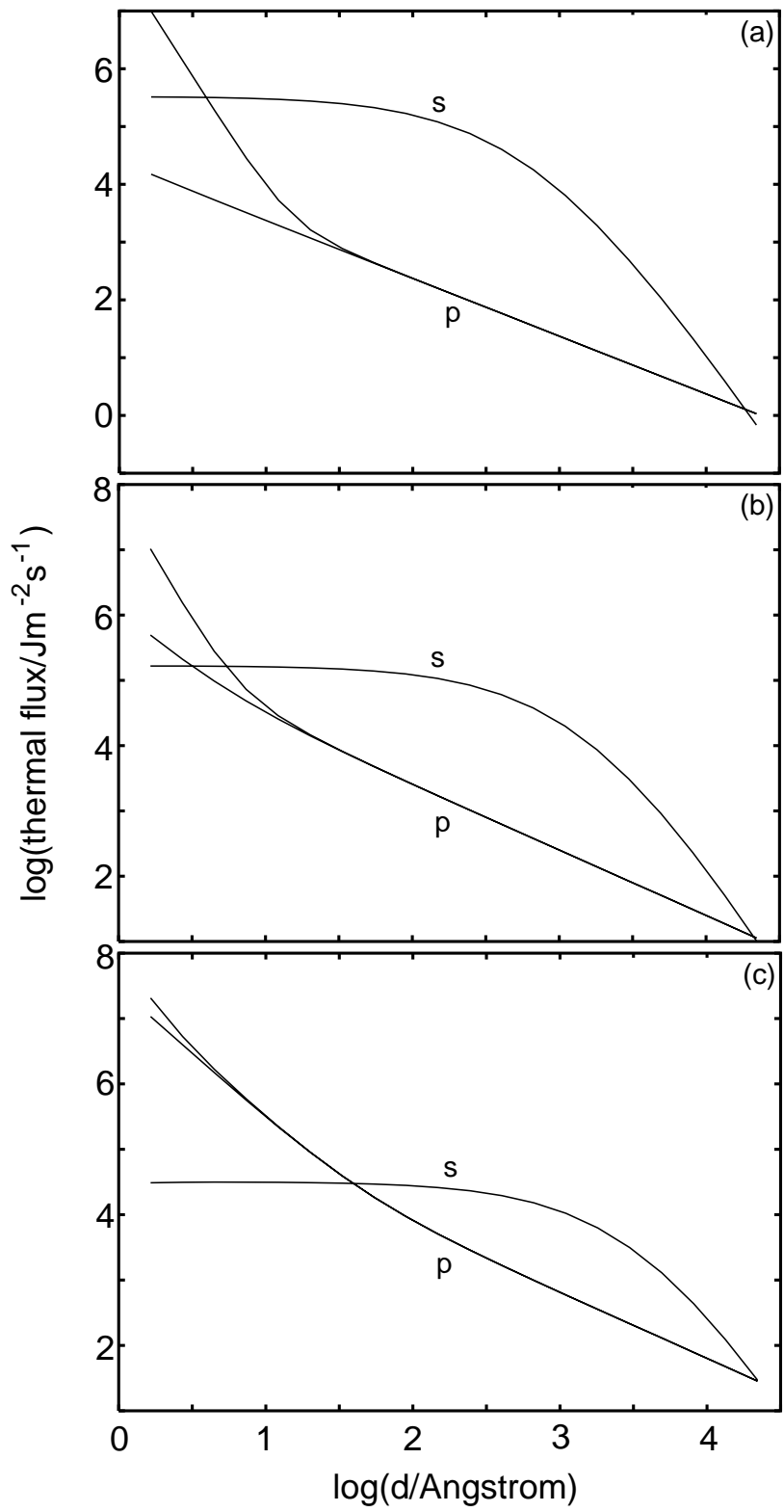
\end{document}